\documentclass[twocolumn,showpacs,preprintnumbers,amsmath,amssymb,superscriptaddress]{revtex4}

\usepackage{graphicx}
\usepackage{dcolumn}
\usepackage{bm}
\usepackage{longtable}

\begin{document}

\preprint{APS/123-QED}

\title{Effective target arrangement in a deterministic scale-free graph}

\author{E. Agliari}
\affiliation{Dipartimento di Fisica, Universit\`a degli Studi di
Parma, viale G.P. Usberti 7/A, 43100 Parma, Italy}
\affiliation{INFN, Gruppo Collegato di
Parma, viale G.P. Usberti 7/A, 43100 Parma, Italy}
\author{R. Burioni}
\affiliation{Dipartimento di Fisica, Universit\`a degli Studi di
Parma, viale G.P. Usberti 7/A, 43100 Parma, Italy}
\affiliation{INFN, Gruppo Collegato di
Parma, viale G.P. Usberti 7/A, 43100 Parma, Italy}
\author{A. Manzotti}
\affiliation{Dipartimento di Fisica, Universit\`a degli Studi di
Parma, viale G.P. Usberti 7/A, 43100 Parma, Italy}

\date{\today}

\begin{abstract}
We study the random walk problem on a deterministic
scale-free network, in the presence of a set of static, identical targets; due to the strong inhomogeneity of the underlying structure the mean first-passage time (MFPT), meant as a measure of transport efficiency, is expected to depend sensitively on the position of targets. We consider several spatial arrangements for targets and we calculate, mainly rigorously, the related MFPT, where the average is taken over all possible starting points and over all possible paths. For all the cases studied, the MFPT asymptotically scales like $\sim N^{\theta}$, being $N$ the volume of the substrate and $\theta$ ranging from $1 - \log 2 / \log3$, for central target(s), to $1$, for a single peripheral target.  
\end{abstract}

 \maketitle

\section{Introduction}
The importance of first-passage phenomena stems from their fundamental role in stochastic processes that are prompted by a first-passage event; examples range from diffusion-limited growth, to neuron firing, and to the triggering of stock options \cite{redner}. These problems can be successfully dealt with by mapping the problem into a random-walk (RW) moving on a proper substrate and in the presence of an absorbing boundary, in such a way that first-passage properties can be expressed in terms of survival probability and mean hitting time. The relation between the first-passage probability and the probability distribution for a RW has been derived in different frameworks (continuous and discrete space and/or time), see e.g.\cite{redner,montroll,weiss,fisher}. In this context, the mean first-passage time (MFPT) to a target point, possibly averaged over all possible starting points, has been widely used in order to measure the transport efficiency \cite{chem,zhang2}.
 
The case of  regular \cite{montroll} and fractal lattices \cite{condamin,olivier2} as substrate has been throughoutly investigated, and recently complex structures, especially (pseudo) scale-free networks, both random and deterministic, have also attracted a lot of interest \cite{bollt,nostro,zhang3,zhang}. The reason is that such networks are able to model some features typical of many real systems and they can exhibit interesting, non-intuitive first-passage properties, whose understanding could pave the way to engineered devices (see e.g.~\cite{cantu,cassi}). In particular, several examples were  shown where the MFPT to the most connected node displays a sublinear dependence on the size $N$ \cite{zhang3,agliari}.

While previous works dealing with complex structures mainly focused on the case of one single trap placed on a central position, here we consider a special realization of deterministic scale-free structure where we study the MFPT for different arrangements of targets. Our results, mainly rigorous, allow to evidence how deeply the position and/or the number of target(s) affect the first-passage properties. Indeed non-trivial effects are found especially when considering large and infinite graphs, which are used to describe macroscopic systems in the thermodynamic limit. More precisely, for the cases analyzed we find that the the MFPT scales like $\sim N^{\theta}$ (up to logarithmic corrections), where $N$ is the number of nodes and $\theta$ is a proper exponent. Moreover, $\theta$ ranges from $\theta_0 = 1 - \log 2/ \log 3 \approx 0.37$ (when one single target is placed on the central node) to $1$ (when the target is placed on a peripheral node), which is expected to be the upper bound \cite{olivier}. Interestingly, the former sublinear scaling is recovered also when a number $N^{\log 2 / \log3}$ of targets is ``centrally arranged''. On the other hand, such an effective performance is by far not representative of the whole network and the MFPT averaged over the target site, denoted as $\bar{\tau}$, is found to scale linearly with $N$. 

The paper is organized as follows: in Sec.~\ref{sec:graph} and Sec.~\ref{sec:VH} we briefly resume the topological properties of the graph under study, while in Sec.~\ref{sec:total} we analyze the average time $\bar{\tau}$. Then, we consider special arrangements of traps distinguishing between ``central'', in Sec.~\ref{sec:effective}, and  ``peripheral'', in Sec.~\ref{sec:ineffective}. Finally, Sec.~\ref{sec:conclusions} contains conclusions and discussions, while in Appendix ~\ref{sec:appe} we have collected the details of analytical calculations. 

\section{The deterministic scale-free graph}\label{sec:graph}
\begin{figure}[h!] 
\begin{center}
 \includegraphics[width=0.45\textwidth]{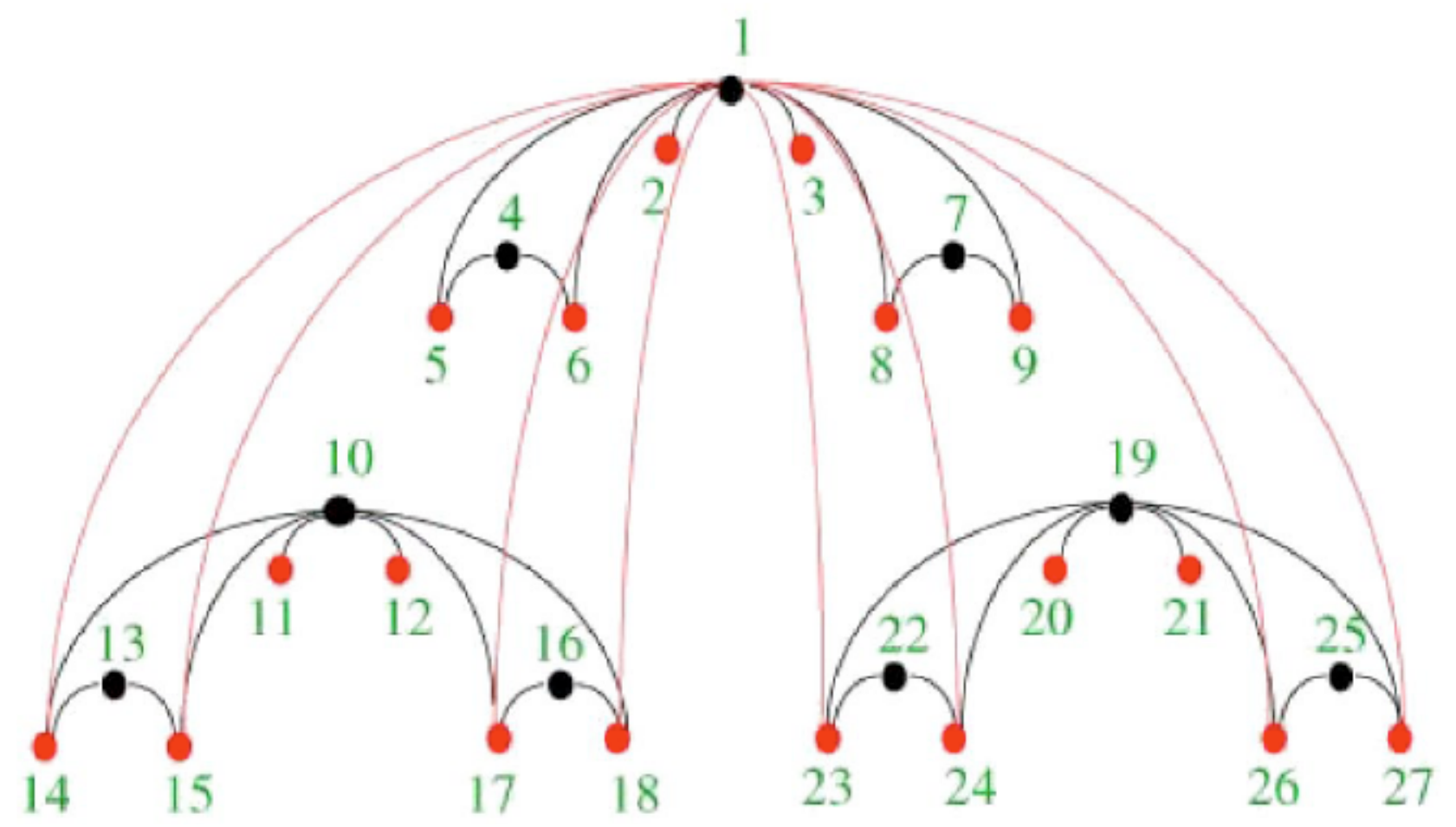}
 \caption{(Color online) Deterministic scale-free graph of generation $g=3$. According to the labeling used $\mathcal{B}_3 = \{ 14, 15, 17, 18, 23, 24, 26, 27 \}$. Moreover, the main hub has label $1$, while hubs of generation $2$ have label $10$ and $19$; analogously, nodes $11,12,20,21$ are rims of generation $1$.}
  \label{fig:grafo}
\end{center}
\end{figure}

The graph under study can be built recursively in such a way that at the $g$th iteration we have the graph of generation $g$, denoted as  $\mathcal{G}_g$ (see Fig.~\ref{fig:grafo} for $g=3$); here we briefly describe its main topological properties, for more details we refer to \cite{barabasi,iguchi,nostro,cinesi_dist}.

Starting from the so-called root constituted by one single node, at the first iteration one adds two more nodes and connects each of them to the root; the resulting chain of length three represents the graph of generation $g=1$. We call $\mathcal{B}_1$ the set of sites introduced at the first generation and linked to the root. Then, at the second iteration one adds two copies of $\mathcal{G}_1$, namely $\mathcal{G}'_1$ and $\mathcal{G}''_1$; for each of them one can distinguish the sets $\mathcal{B}'_1$ and $\mathcal{B}''_1$, which make up the set $\mathcal{B}_2$, i.e. $\mathcal{B}_2 = \mathcal{B}''_1 \cup \mathcal{B}''_1$; each node in $\mathcal{B}_2$ is then directly connected to the root. Proceeding analogously, at the $g$-th iteration one introduces two replica of the existing graph of generation $g-1$ and connects the root with all the sites belonging to $\mathcal{B}_g$, henceforth called ``rims''.

This way, at the $g$-th generation the total number of nodes
is $N_g=3^g$, the total number of rims is
$|\mathcal{B}_g|=2^g$ and the root turns out to be the main
hub with coordination number $2(2^{g}-1)$. Apart from the main hub, one can detect ``minor hubs'', namely the two hubs pertaining to the subgraphs $\mathcal{G}'_{g-1}$ and 
$\mathcal{G}''_{g-1}$; more generally, one can detect $2^2$ so-called hubs of generation $g-2$ pertaining to the two copies of $\mathcal{G}'_{g-2}$ and $\mathcal{G}''_{g-2}$ and so on in cascade. Analogously, we will refer to rims of generation $g-n$, with $n=0,1,...,g-1$ (see Fig.~\ref{fig:grafo}). 
Now, hubs follow a power-law degree distribution: $P(k) \sim k^{-\gamma}$ with
exponent $\gamma=\log 3 / \log2 \approx 1.59$ \cite{barabasi}, while the remaining nodes follow an exponential degree distribution: $P(k) \sim (2/3)^k =
e^{-\bar{\gamma} k}$, where $\bar{\gamma} = \log(3/2)\approx 0.405
$ \cite{iguchi}. 

Finally, the average degree is $\langle k \rangle_g \equiv \sum k P(k) / N_g = 4 [1-(2/3)^g]$,
namely it approaches $4$ as $g \rightarrow \infty$;
on the other hand, the second moment $\langle k^2 \rangle$ is divergent \cite{iguchi}.

\section{\label{sec:VH} Van Hove sphere and average distances}
For a graph of generation $g$, being $X(k,g)$ the number of nodes at a distance $k$ from the main hub and, analogously,  $Y(k,g)$ the number of nodes at a distance $k$ from an arbitrary rim, the following equations hold:
\begin{eqnarray}
\label{eq:X}
X(k,g) = X(k,g-1)+2Y(k-1,g-1) \\
\label{eq:Y}
Y(k,g)=X(k-1,g-1)+2Y(k,g-1).
\end{eqnarray}

Notice that $X(k,g)$ represents the cardinality of the borders of the generalized Van Hove sphere with radius $k$ in the chemical distance and centered in the main hub \cite{cassi}.
Passing to the corresponding generating functions, namely $\tilde{X}(s,g) \equiv \sum_{k=0}^{\infty} X(k,g) s^g$ and $\tilde{Y}(s,g) \equiv \sum_{k=0}^{\infty} Y(k,g) s^g$, we get
\begin{eqnarray}
\label{eq:Xg}
\tilde{X}(s,g)= \tilde{X}(s,g-1)+2 \tilde{Y}(s,g-1) s\\
\label{eq:Yg}
\tilde{Y}(s,g)=\tilde{X}(s,g-1)s+2 \tilde{Y}(s,g-1).
\end{eqnarray}
Now, by combining Eq.~\ref{eq:Xg} and Eq.~\ref{eq:Yg} we have
\begin{equation}
\label{eq:X}
\tilde{X}(s,g+2)=3 \tilde{X}(s,g+1)-2(1-s^2) \tilde{X}(s,g).
\end{equation}
The solution of this finite difference equation can be obtained by recalling the initial condition
\begin{eqnarray}
X(k,0) =\delta_{k,0} \; \Rightarrow \tilde{X}(s,0)=1 \\
X(k,1)= \delta_{k,0} + \delta_{k,1}  \; \Rightarrow \tilde{X}(s,1)=1+2s,
\end{eqnarray}
due to the fact that the root of the graph (generation zero) is just the main hub, while the first generation graph is made up by the main hub and by two rims only.
Hence we get \cite{prox}
\begin{eqnarray}
\nonumber
X(2k,g) &=&  \left( \frac{3}{2} \right)^g 2^{3k} \\
\label{eq:even}
&\times&  \sum_{l=0}^{g/2}\frac{1}{3^{2l}} \binom{l}{k}  \binom{g}{2l}  \left[ 1 - \frac{g-2l}{3(2l+1)} \right] ,
\end{eqnarray}
\begin{equation}
\label{eq:odd}
X(2k+1,g) =  \left( \frac{3}{2} \right)^g \frac{4}{3} \; 2^{3k} \sum_{l=0}^{g/2} \frac{1}{3^{2l}} \binom{l}{k}  
\binom{g}{2l+1},  
\end{equation}
which provide the cardinality of the van Hove surface of even and odd radius, respectively. Interestingly,
these expressions clarify the peculiar topology of the deterministic scale free graph: due to the 
non monotonicity of $X(k,g)$ and to the large number of surface points at a given distance, quantities obtained  from the limit of a sequence of larger and larger  graphs  could present some anomalies \cite{prox}.

From $\tilde{X}(s,g)$ we can also derive the average distance from 
the main hub $d_H$.
Let us define $d_{ij}$ the average (chemical) distance from $i$ to $j$, then we have
\begin{equation}
d_g = \frac{\sum_{i \in N_g} d_{iH}}{N_g},
\end{equation}
where $H$ is the node representing the main hub.
Now, the following relation holds
\begin{equation}
d_g = \frac{1}{N_g} \frac{\partial}{\partial s} \tilde{X}(s,g) \Bigg |_{s=1}.
\end{equation}
Then, from Eq.~\ref{eq:X}, we get 
\begin{eqnarray} \label{eq:recurrence}
\nonumber
\frac{\partial}{\partial s} \tilde{X}(s,g+2) &=& 3  \frac{\partial}{\partial s} \tilde{X}(s,g+1) + 4 s \tilde{X}(s,g) \\ 
&-& 2(1-s^2) \frac{\partial}{\partial s} \tilde{X}(s,g),
\end{eqnarray}
which, for $s=1$, gives 
\begin{equation} \label{eq:recurrence}
N_{g+2} \, d_H (g+2) = 3 N_{g+1} \, d_H(g+1) + 4 \tilde{X}(1,g) = 0.
\end{equation}
From Eq.~\ref{eq:X}, $\tilde{X}(1,g) = 3^g = N_g$ follows, as expected, and by plugging $\tilde{X}(1,g)$ into Eq.~\ref{eq:recurrence}, we obtain the following recursive relation
\begin{equation} \label{eq:recurrence2}
9 d_{g+2} = 9 d_{g+1} + 4= 0,
\end{equation}
whose solution is
\begin{equation} \label{eq:distance}
 d_g = \frac{2}{9} (2 g + 1),
\end{equation}
where we used the initial condition $d_1 = 2/3$.
The average distance to the main hub can also be expressed as a function of the volume, recalling that $g = \log N_g / \log 3$
\begin{equation} \label{eq:distance2}
 d_g = \frac{2}{9} \left( 2 \frac{\log N_g} { \log 3} + 1 \right) \sim \log N_g.
\end{equation}

While $d_g$ provides information about the ``accessibility'' of the main hub itself, the average distance $\bar{d}_g$ calculated over all pairs of nodes, provides a more global information, being tightly connected with dynamics phenomena on graphs such as spreading, random walks and synchronization \cite{watts,condamin,motter}.  The average distance is defined as
\begin{equation}
\bar{d}_g \equiv \frac{\sum_{i,j=1}^{N_g} d_{ij} }{N_g(N_g -1)}.
\end{equation}
The exact expression of $\bar{d}_g$ for the graph under study has been obtained in \cite{cinesi_dist} and is given by
\begin{equation}\label{eq:dist_media}
\bar{d}_g = \frac{ 8 N_g \log N_g }{9 (N_g -1) \log 3} \sim \log N_g,
\end{equation}
where the last expression holds in the thermodynamic limit $N_g \to \infty$.

The consistency between the leading behaviors of $d_g$ and $\bar{d}_g$, see Eqs.~\ref{eq:distance2} and \ref{eq:dist_media}, suggests that, despite its centrality, the main hub is not, in the average, spatially closer than any other site. In fact, the maximum distance among nodes grows logarithmically with the volume, that is, $\max_{i,j=1,...,N_g} d_{ij} = 2g$, and the graph is generally rather compact.

\section{Total mean first passage time}\label{sec:total}

The average distance $\bar{d}_g$ discussed in the previous section allows to determine an upper bound for the total mean first passage time $\bar{\tau}_g$ defined as  
\begin{equation} \label{eq:total_tau}
\bar{\tau}_g  \equiv \frac{1}{N_g(N_g-1)}\sum_{i\neq j}^{N_g} \sum_{j=1}^{N_g} t_{ij},
\end{equation}
where $t_{ij}$ is the mean time taken by a simple RW on $\mathcal{G}_g$ to first reach the site $j$ starting from $i$, and the average is taken over all possible couples. As shown in \cite{olivier,olivier3}, $\bar{\tau}_g$ is subject to the following, rather strict, constraints:
\begin{equation}\label{eq:constraints}
\frac{N_g}{2} \leq \bar{\tau}_g \leq \frac{N_g \langle k \rangle \bar{d}_g }{2} \sim N_g \log N_g,
\end{equation}
where we used Eq.~\ref{eq:dist_media} for $\bar{d}_g$ \cite{cinesi_dist}.

Such an average is characteristic of a network, in the sense that it provides an overall measure of the transport efficiency on that network \cite{zhang}, but, on the other hand, it conceals the peculiarities of the graph itself and it does not allow to see whether there exists any target position which results to be particularly effective or ineffective. 

In order to deepen this point we start by considering  the set $\{ t_{ij} \}_{i,j=1,...,N_g}$:
For an arbitrary connected couple $(i,j)$, the mean time $t_{ij}$ can be calculated numerically by exploiting a method based on the so-called pseudoinverse Laplacian $\mathbf{\tilde{L}}^{\dagger}$ \cite{cantu,zhang4}.

\begin{figure}[ht] 
\begin{center}
 \includegraphics[width=0.5\textwidth]{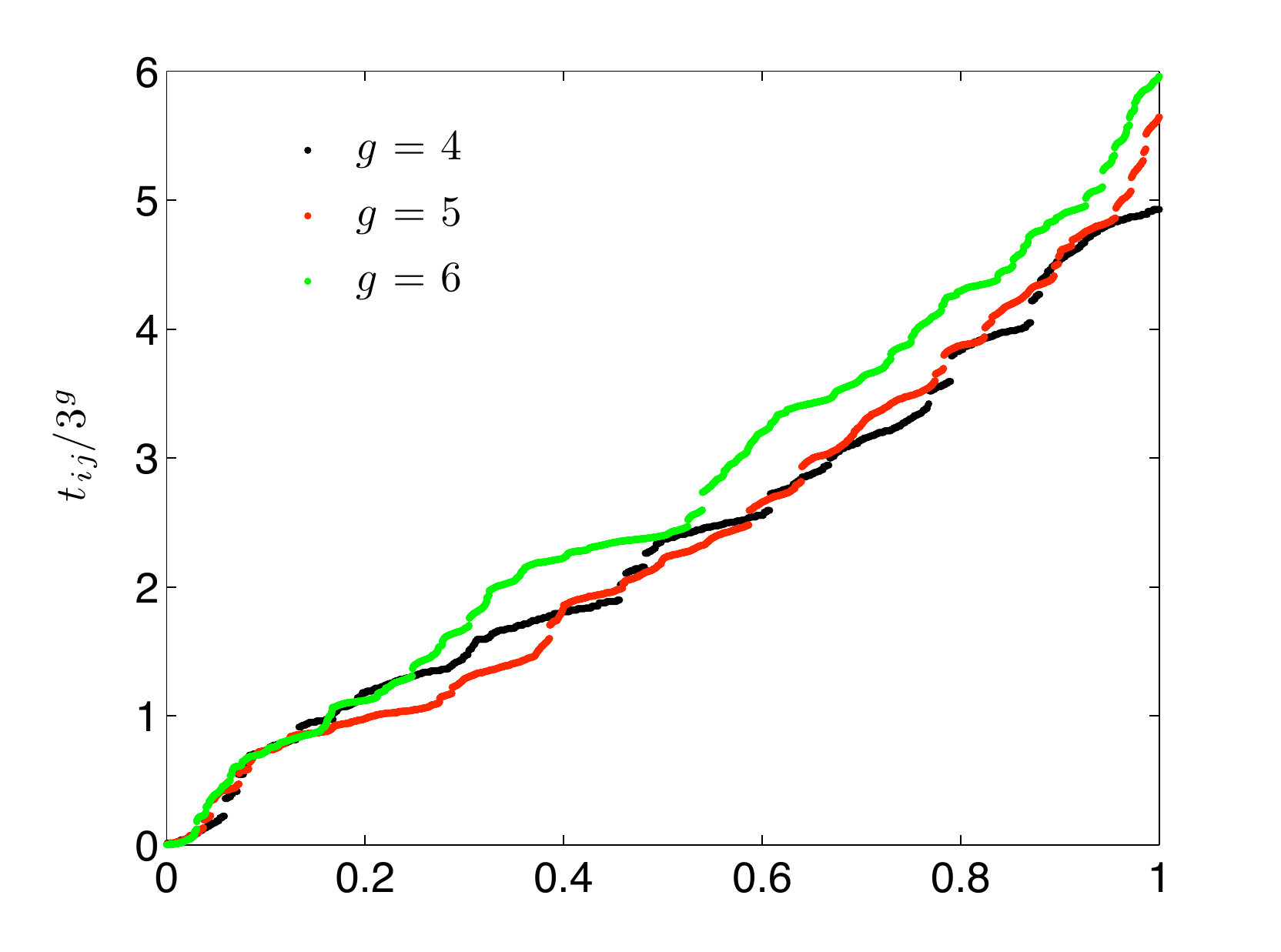}
 \caption{(Color on line) Mean times $t_{ij}$ divided by the volume $N_g$ for graphs of generations $g=4,5,6$, as indicated by the legend; only non degenerate data have been depicted. Notice that data are plotted in ascending order and versus a proper normalized interval.}
  \label{fig:times}
\end{center}
\end{figure}

\begin{figure}\label{cantu_par}
\begin{center}
 \includegraphics[width=0.5\textwidth]{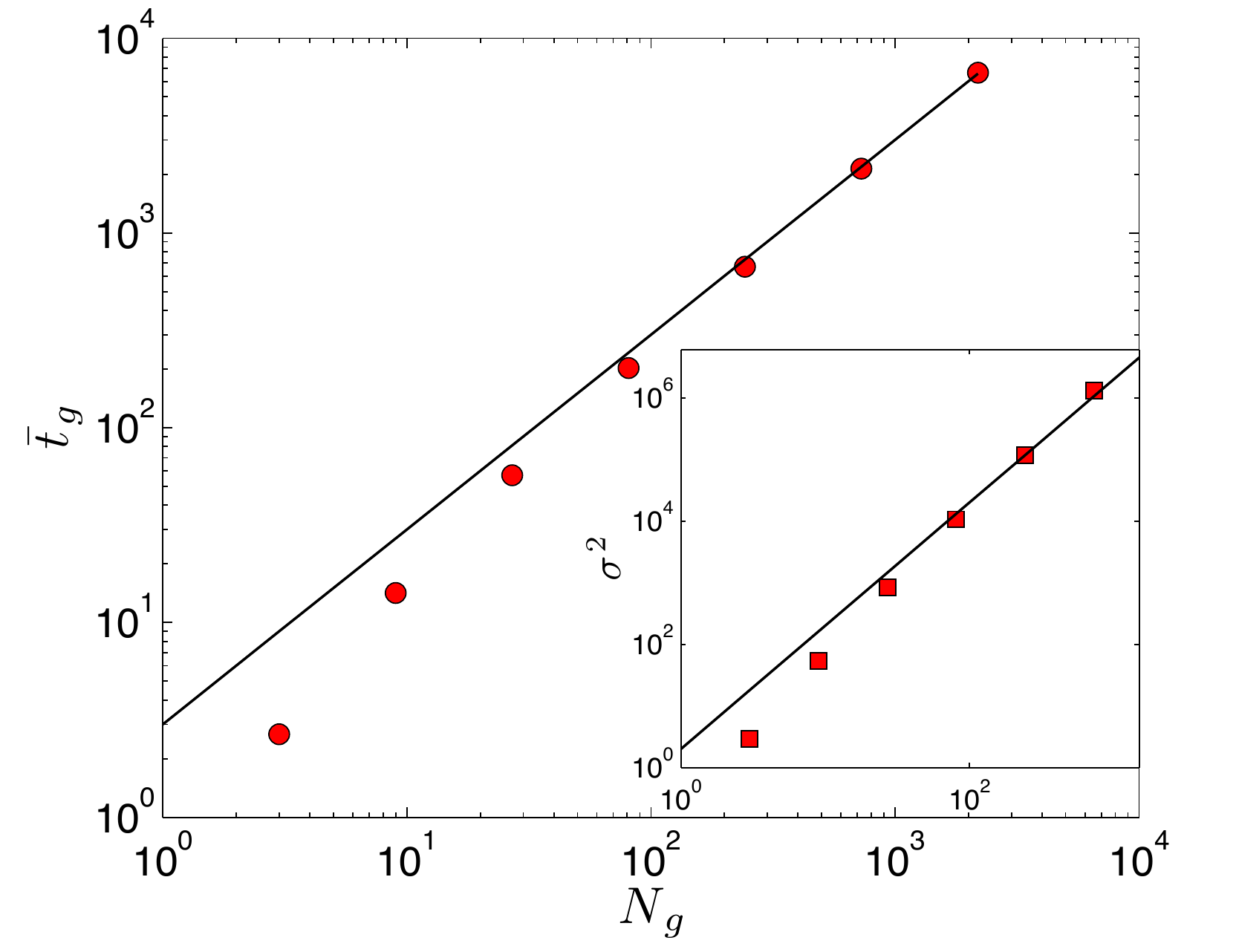}
 \caption{(Color on line) Total mean first passage time $\bar{\tau}_g$ (main figure) and related variance $\sigma^2$ (inset) as a function of the graph volume $N_g$. Data points have been fitted with the power law $\sim N_g$ and $\sim N_g^2$, respectively.}
\end{center}
\end{figure}

The intrinsic symmetry of the graph yields degenerate values for $t_{ij}$ which have been omitted in the plot of Fig.~\ref{fig:times}. From the plot one can notice that $t_{ij}$ spans a wide range $\sim \mathcal{O}(N_{g})$ and the variance $\sigma^2$ of the set $\{ t_{ij} \}$ (see the inset of Fig.~\ref{cantu_par}) is shown to grow asymptotically as the square of the graph volume. The large variability of $t_{ij}$ mirrors the strong inhomogeneity of the graph under study; this is a further confirmation that the efficiency of first-passage processes is sensitively affected by the target position.

From the numerical data collected we are able to calculate $\bar{\tau}_g$ according to Eq.~\ref{eq:total_tau}: as shown in Fig.~\ref{cantu_par} (main figure) its asymptotic behavior is consistent with the linear law $\bar{\tau}_g \sim 3 N_g$, in agreement with Eq.~\ref{eq:constraints}.

\section{Target on central nodes}\label{sec:effective}
As shown in \cite{nostro}, for a simple RW on $\mathcal{G}_g$, the mean time to first reach the main hub starting from any of the rims in $\mathcal{B}_g$ is
\begin{equation}\label{eq:R_g}
R_g= \frac{8}{3} \left( \frac{3}{2} \right)^g - 3, 
\end{equation}
while the mean time to first reach any rim starting from the main hub is
\begin{equation}\label{eq:H_g}
H_g = \frac{4}{3} \left( \frac{3}{2} \right)^g - 1, 
\end{equation}
where the mean is taken over all possible paths. Such results were used to calculate the mean time to first reach the main hub $\tau_g(H)$ (where the mean is taken over all possible paths and over all possible starting points), which reads as 
\begin{equation}\label{eq:main_hub}
\tau_g = \frac{1}{3^g - 1} \left[ \frac{32}{9} \left( \frac{9}{2} \right)^g - \frac{2}{9} (17 + 4g)3^g  \right] \sim N_g^{1 - 1/\gamma},
\end{equation}
being $1- 1/\gamma= \theta_0 = 1 - \log 2 / \log 3 \approx 0.37$.
Notice that the asymptotic behavior of $R_g $, $H_g$ and $\tau_g(H)$ is the same, namely $(3/2)^g$.
The exponent $\theta_0$ is very small if compared with the result found for $\bar{\tau}_g$ and also with similar results found in the literature \cite{kozak,agliari,nostro,zhang3,zhang}, making the main hub a particularly effective node where to place a target.

We now calculate the mean hitting time $\tau_g(\mathcal{B}_g)$ at any rim, averaged over all possible starting points $\notin \mathcal{B}_g$ and over all possible paths:
\begin{equation}
\tau_g(\mathcal{B}_g) \equiv \frac{1}{N_g - |\mathcal{B}_g|} \sum_{i \notin \mathcal{B}_g} t_{ib}.
\end{equation}
By properly combining $H_g $ and $R_g$ we can write the following recursive equation:
\begin{eqnarray}\label{eq:ricorsive}
\nonumber
\tau_g(\mathcal{B}_g) &=& \frac{1}{3^g-2^g}[3^{g-1} H_g+(3^{g-1}-1)\tau_{g-1} \\
&+&2\tau_{g-1}(\mathcal{B}_{g-1}) (3^{g-1}-2^{g-1})].
\end{eqnarray}
In fact, if the RW starts from a site belonging to the subgraph $\mathcal{G}_{g-1}$ containing the main hub, then, in order to arrive to rims, it must pass through the main hub itself (in a mean time $\tau_{g-1}(H)$) from which rims are first reached in a time $H_g$. Conversely, if the starting point is in $\mathcal{G}_{g-1}' \cup \mathcal{G}_{g-1}'' $, the MFPT to rims is just $\tau_{g-1}(\mathcal{B}_{g-1})$.
The recursive relation can be solved to obtain
\begin{equation}\label{eq:somme}
\tau_g(\mathcal{B}_g) = \frac{3^g}{9(3^g-2^g)}\left[ 20 \left(\frac{3}{2} \right)^g -8 g+18 \right] \sim \left(\frac{3}{2} \right)^g.
\end{equation}
Interestingly, the leading behaviour is the same as $\tau_g(H)$, even if in this case the number of targets is $2^g \sim N_g^{1 / \gamma}$.
It is rather intuitive at this point that when targets are placed on the main hub \emph{and} on rims the leading behaviour of the hitting time is conserved. Indeed, one has
\begin{eqnarray}\label{final:HR}
\nonumber
\tau_{g}(H \cup \mathcal{B}_g)&=&\frac{1}{3^{g}-2^{g}-1}[ (3^{g-1}-1)\tau_{g-1}\\ 
&+& \tau_{g-1}(\mathcal{B}_{g-1})(2\cdot3^{g-1}-2^{g}) ].
\end{eqnarray}

\begin{figure}[htb] 
  \centering
    \includegraphics[width=0.5\textwidth]{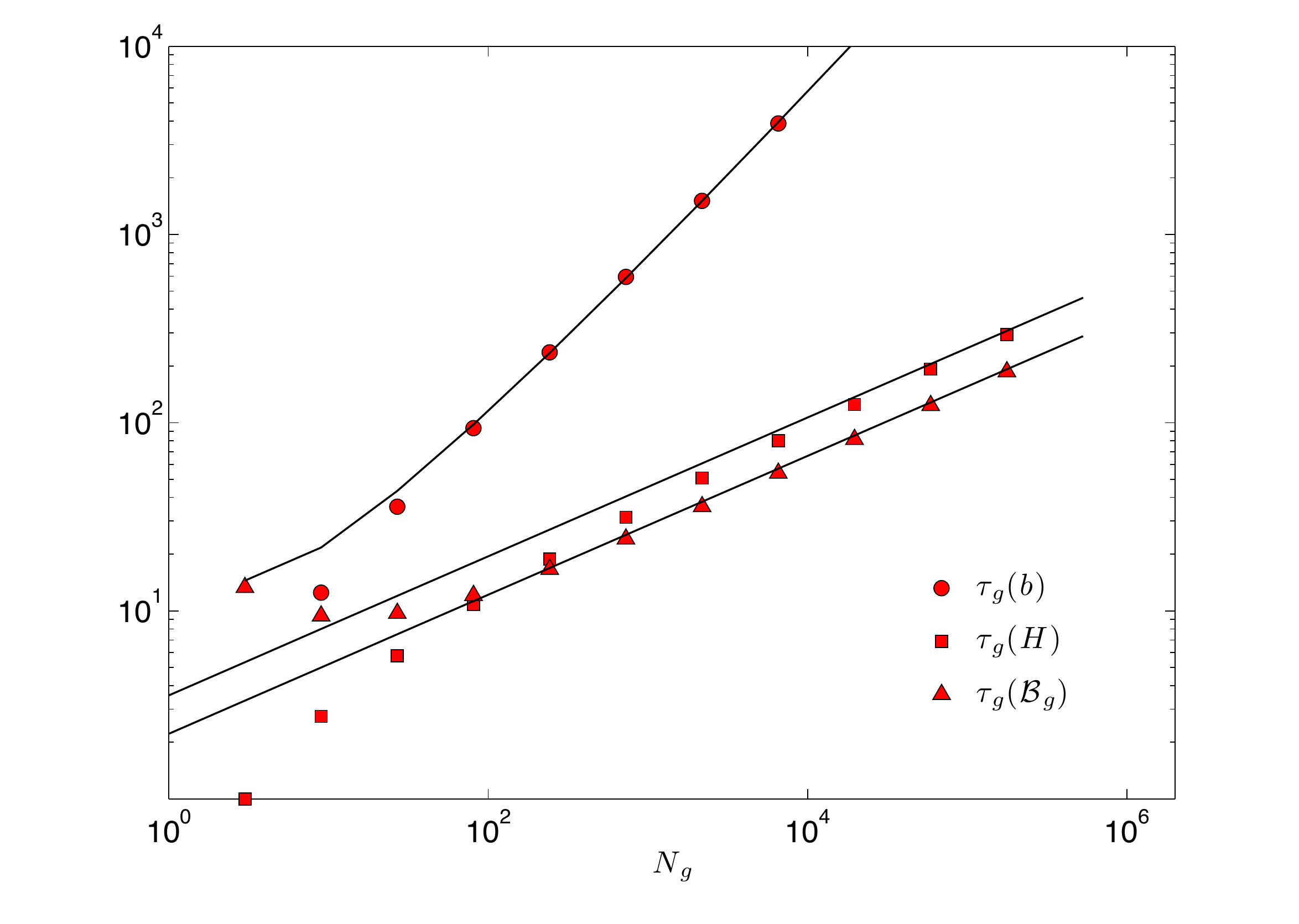}
     \caption{(Color on line) Mean hitting times at the main hub $\tau_g(H)$, at rims $\tau_{g}(\mathcal{B}_g)$ and at one rim $\tau_{g}(b)$, as a function of the number of nodes $N_{g}$ for a simple RW moving on $\mathcal{G}_g$. Data from numerical calculations coincide with those calculated analytically (for the former to cases see Eq.~\ref{eq:main_hub}, Eq.~\ref{eq:somme}) and are compared with the leading behaviors $\sim N_g^{1-1/\gamma}$ and $\sim N_g/ \log N_g$ (continuous line).}
     \label{fig:rim}
\end{figure}

We finally consider the case of one single target placed on a rim, i.e. on a site $b \in \mathcal{B}_g$; the calculation has been performed numerically exploiting the following expression \cite{kozak}
\begin{equation}
\nonumber
\tau_g (b) = \frac{1}{N_g - 1} \sum_{\substack{i =1 \\ i \neq b}}^{N_g} t_{ib} = \frac{1}{N_g - 1} \sum_{\substack{i =1\\ i \neq b}}^{N_g} \sum_{\substack{j = 1\\ j \neq b}}^{N_g}  (-\mathbf{L}^{-1})_{ij},
\end{equation}
where $\mathbf{L}$ is the Laplacian matrix of the underlying structure.
In the range considered, i.e. $g \in [0,6]$, data points can be fitted by the following
\begin{equation}
\tau_{g}(b) \sim \frac{N_g}{\log N_g},
\end{equation}
which is consistent with the analytical result found in \cite{olivier} (although the average here is taken over the flat distribution).
The MFPT's $\tau_g(H)$, $\tau_{g}(\mathcal{B}_g)$ and $\tau_{g}(b)$ are shown and compared in Fig.~\ref{fig:rim}

\section{Target on peripheral nodes}\label{sec:ineffective}
After having analyzed particularly effective arrangements for targets, i.e. main hub and rims, we now want to focus on trapping at the farthest nodes from the main hub which are expected to be rather difficult to reach.

We notice that in a graph of generation $g$, the radius of the outmost van Hove surface centered on the main hub, i.e. the largest chemical distance achievable from the main hub, equals $g$ and its cardinality is $X(g,g) = g$ if $g$ is even, otherwise $X(g,g) = g+1$; such farthest nodes correspond to hubs ($g$ even) or rims ($g$ odd), of subgraphs of generation $1$. 
We call $\mathcal{F}_g$ the set of nodes at a chemical distance $g$ from the main hub so that $|\mathcal{F}_g| = X(g,g)$.

We will first consider the case of $\mathcal{F}_g$ working as a boundary, that is each node in $\mathcal{F}_g$ is occupied by a target, hence, exploiting the intrinsic symmetry of the arrangement, the problem will be solved analytically. Later we will focus on the case of one single target fixed on a node belonging to $\mathcal{F}_g$ and we will solve the problem numerically.

\subsection{Multi-target}
The basic step to calculate the hitting time $\tau_g(\mathcal{F}_g)$ when each node in 
$\mathcal{F}_g$ is occupied by a target, is to obtain $H_{g-n}$ and $R_{g-n}$, which represent, respectively, the MFPT from the hub of a subgraph of generation $g-n$ to any rim of the same subgraph, and the MFPT from any rim in a subgraph of generation $g-n$ to the pertaining hub; for example, referring to Fig.~\ref{fig:grafo}, 
$R_{3-1}$ is the mean time to go from node $14$ to $10$, while $H_{3-2}$ is the mean time to reach $11$ starting from $10$ (and equivalently for similar nodes).
The detailed calculations are reported in the Appendix, while here we just summarize the main passages and results.

We first consider the MFPT from the main hub to $\mathcal{F}_g$; starting from the main hub the RW has first to reach the rims in a time $H_g$ (see Eq.~\ref{eq:H_g}), from $\mathcal{B}_g$ the RW has to reach the hub of the subgraph of generation $g-1$ and this is found to be
\begin{equation}
R_{g-1}=\frac{28}{3}\left(\frac{3}{2}\right)^{g}-3,
\end{equation}
which exhibits the same asymptotic behavior of $H_g$ and of $R_g$ (see Eqs.~\ref{eq:R_g} and \ref{eq:H_g}) \cite{nostro}.
Then, the RW has to ``bounce'' from rim to hub and from hub to rims of inner and inner subgraphs, the pertaining MFPT's read as
\begin{equation} \label{eq:rimbalzo1}
R_{g-n}=4 \times 2^{n/2} \left[ \left(\frac{3}{2}\right)^{g} -1 \right] +1 -\frac{8}{9}\left(\frac{3}{2}\right)^{g-n}
\end{equation} 
and
\begin{equation} \label{eq:rimbalzo2}
H_{g-n}=4 \times 2^{n/2} \left[ \left(\frac{3}{2}\right)^{g} -1 \right] +3 -\frac{8}{3}\left(\frac{3}{2}\right)^{g-n}.
\end{equation}
We notice that $R_{g-n}$ and $H_{g-n}$ (Eqs.~\ref{eq:rimbalzo1} and \ref{eq:rimbalzo2}) display the same asymptotic behavior, namely 
$(3/2)^g 2^{n/2}$, which, interestingly, is the same behavior found for $\tau_g(H)$ but ``corrected'' by a factor which depends on $n$ as $2^{n/2}$:
\begin{eqnarray}
R_{g-n} \sim 2^{n/2} \; R_g \\
H_{g-n} \sim 2^{n/2} \; H_g.
\end{eqnarray}
Some remarks are in order here. Although, for a given subgraph, the distance between the hub and a rim is the same and equal to $1$, the mean time taken by the RW depends significantly on how ``deep'' the subgraph is (i.e. how small $g-n$ is). Such a result is somehow counterintuitive because the probability to go directly from rim to hub $[1/(g-n+1)]$ and from hub to rim $[2^{g-n-1}/(2^{g-n}-1)]$ grows with $n$. Actually, what really matters is that if the right way is missed at the first step then it is more and more difficult for the RW to retrieve it as $n$ gets larger. The factor $2^{n/2}$ is a clear indication of the fact that targets placed on inner nodes are in general rather difficult to reach.

\begin{figure}[h!] 
  \centering
    \includegraphics[width=0.5\textwidth]{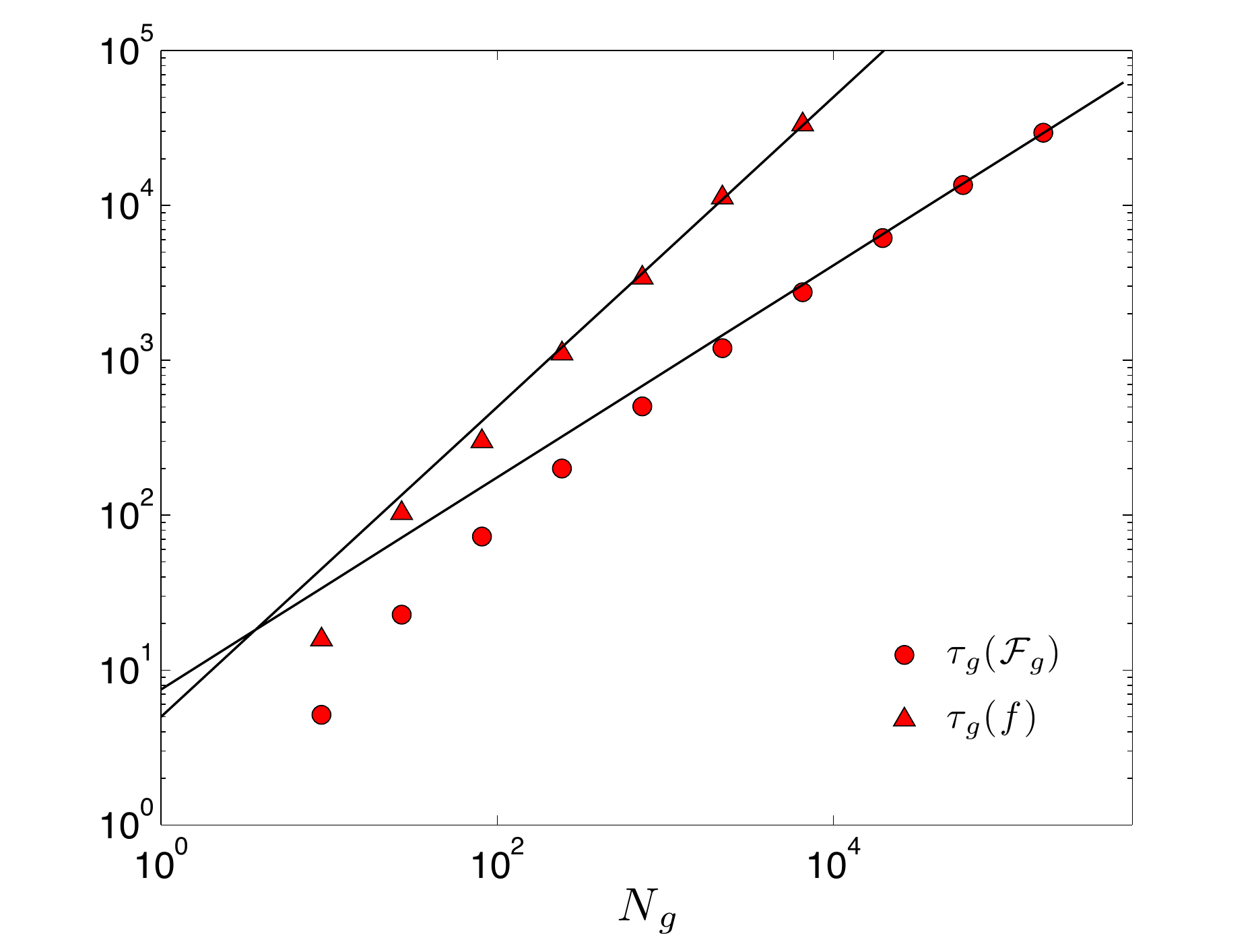}
     \caption{(Color on line) Mean hitting times at peripheral nodes $\tau_{g}(\mathcal{F}_g)$ ($\bullet$) and $\tau_{g}(f)$ ($\blacktriangle$), as a function of the number of nodes $N_{g}$ for a simple RW moving on $\mathcal{G}_g$. Data from numerical calculations coincide with those calculated analytically from Eq.~\ref{eq:somme} and are compared with the leading behavior $\sim N_g^{1-1/(2\gamma)}$ and $\sim N_g$ (continuous line), according to Eq.~\ref{eq:uno} and Eq.~\ref{eq:due}, respectively.}
     \label{fig:one}
\end{figure}

By merging the previous results one can find the average time $t_{g-n}^H$ to reach a node in $\mathcal{F}_g$, starting from the hub of any subgraph of generation $g-n$
\begin{equation} \label{tempo_h}
t_{g-n}^H(\mathcal{F}_g)=\sum_{l}^{g-2}(R_{g-l}+H_{g-l}) 
\end{equation}
or from any rim of generation $g-n$
 \begin{equation} \label{tempo_h2}
t_{g-n}^R(\mathcal{F}_g)=\sum_{l-1}^{g-2}R_{g-l}+\sum_{l=g-n-1}^{g-2}H_{g-l}.
\end{equation} 
These expressions are used in Appendix \ref{sec:appe} to get the average time $\tau_g(\mathcal{F}_g)$ to reach an arbitrary node $f \in \mathcal{F}_g$, where the average is meant over all possible starting points $\notin \mathcal{F}_g$ and over all possible paths, namely
\begin{equation}
\tau_g(\mathcal{F}_g) \equiv \frac{1}{N_g - |\mathcal{F}_g|} \sum_{i \notin \mathcal{F}_g} t_{if}.
\end{equation}
Hence, we finally get
\begin{eqnarray}\label{final}
\nonumber
\tau_g(\mathcal{F}_g)&=&\frac{2^{g+1}}{45(3^g - 2^{\frac{g}{2}})} \Big[ 4 \left( \frac{3}{2} \right)^{2g} (45 \times 2^{g/2} - 97)\\ 
\nonumber
&-&   \left( \frac{3}{2} \right)^{g} (45 \times 2^{g/2+2}  -25 g - 527)  -144 \Big]\\
&\sim&  \left( \frac{3}{2} \right)^{g}  2^{g/2} \sim N_g^{1-\frac{1}{2 \gamma}},
\end{eqnarray}
where the last line contains the asymptotic behavior valid in the limit $N_g \to \infty$ (see Fig.\ref{fig:one}).
Although the number of targets is $\sim \log N_g$, the mean time to first reach any of them grows much faster than the mean time to reach the main hub $\tau_g(H) \sim (3/2)^g$, but still slower than the average hitting time $\bar{\tau}_g \sim 3^g$. Indeed, recalling $1-\frac{1}{2 \gamma} = 1-\frac{1}{2} \frac{\log 2}{\log 3} \approx 0.6845$, we have that 
\begin{equation}\label{eq:uno}
\tau_g(\mathcal{F}_g)  \sim  \bar{\tau}_g^{1-\frac{1}{2 \gamma}} \sim \tau_g(H)^{1 + \frac{1}{2(\gamma -1)}},
\end{equation}
where $1 + 1/(2 \gamma -2) \approx 1.85$.

\subsection{Single Target}
We now turn to the case of one single target placed on a node $f \in \mathcal{F}_g$, which has been investigated numerically according to
\begin{equation}
\nonumber
\tau_g (f) = \frac{1}{N_g - 1} \sum_{\substack{i =1 \\ i \neq f}}^{N_g} t_{if} = \frac{1}{N_g - 1} \sum_{\substack{i =1\\ i \neq f}}^{N_g} \sum_{\substack{j = 1\\ j \neq f}}^{N_g}  (-\mathbf{L}^{-1})_{ij}.
\end{equation}
Results found for several generations are shown in Fig.\ref{fig:one}.
By means of fitting procedures we find that when the size is large, the mean hitting time $\tau_g (f)$ scales like
\begin{equation}\label{eq:due}
\tau_g (f) \sim N_g^{\theta_f},
\end{equation}
where the exponent $\theta_f$ is consistent with the value $1$, in agreement with \cite{olivier}.

\section{Conclusions} \label{sec:conclusions}
In this paper we studied the mean first-passage problem for a random walk on a deterministic scale-free structure, focusing the attention on different arrangement of targets. First of all, we highlighted that, due to the high inhomogeneity of the structure, the mean first-passage time (MFPT) $t_{ij}$ from a node $i$ to $j$ sensitively depends on the couple $(i,j)$ chosen, spanning a range $\sim \mathcal{O}(N_g)$, being $N_g$ the size of the structure.  

The case of a single target placed on the main hub turns out to be strategic as it gives rise to a slow sublinear scaling $\sim N_g^{\theta_0}$, being $\theta_0 = 1 - 1 / \gamma = 1 - \log2 / \log 3 \approx 0.37$. Interestingly, the same scaling is obtained in the case that the whole set of rims $\mathcal{B}_g$, with cardinality $\sim \mathcal{O}(N_g^{1 / \gamma})$, is assumed as absorbing boundary. On the other hand, when only one single node $b \in \mathcal{B}_g$ works as a target we get a faster growth $\sim N_g / \log N_g$, where the logarithmic correction accounts for the large coordination number of rims (see \cite{olivier}).

When peripheral positions for traps are considered, sublinear scaling are achievable provided that the number of targets is large enough; in particular, when the absorbing boundary is given by the whole set of peripheral nodes $\mathcal{F}_g$ with cardinality $\log N_g / \log 3$ we have that the mean hitting time scale like $\sim N_g^{1 - 1/(2 \gamma) }$. Conversely, when one single node $f \in \mathcal{B}_g$ works as a target our fits suggest a linear growth $\sim N_g$.

Therefore, the transport efficiency sensitively depends on the number and on the position of targets, moreover, the existence of very ``close'' couples, such as main hub and rims, is evened off by rather ``distant'' couples, giving rise to a mean hitting time $\bar{\tau}_g$, averaged over all possible pairs, which scales linearly with $N_g$.

\appendix
\section{Mean FPT on $\mathcal{F}_g$}\label{sec:appe}
In this appendix we report the detailed calculation of the mean time taken by a simple RW on a graph $\mathcal{G}_g$ and started from the hub of a subgraph of generation $n$, to first reach any rim in the same subgraph, which is denoted as $H_{n}$, and the MFPT from any rim in a subgraph of generation $n$ to the pertaining hub, denoted as $R_{n}$.
Such times will then be merged in order to obtain the mean time $\tau_g(\mathcal{F}_g)$ (averaged over all possible starting points and over all possible paths) to first reach any node belonging to the set $\mathcal{F}_g$ and meant to be occupied by a target.
 
 Let us start from the highest generation, i.e. $n=g$.
As for $H_g$, this has already been calculated in \cite{nostro} (see Eq.~\ref{eq:H_g}).
Now, from rims of generation $g$, in order to reach $\mathcal{F}_g$, we first need to pass through the hub of gneration $g-1$ and we can write
\begin{eqnarray}\label{eq:rhprima}
R_{g-1}=\frac{1}{g} \Big[ (t_{g}^{H}+1+ R_{g})+1+\sum_{l=1}^{g-2} (1+t_{l}^{H}+ R_{g}) \Big].
\end{eqnarray}
In fact, starting from any rim of generation $g$, a RW has $g$ options: it goes to the main hub and then returns to rims in a time $H_g$, either it goes directly to the hub of generation $g-1$, either it reaches a hub of generation $g-l$ from which it return to rims in an average time $H_l$. By plugging Eqs.~\ref{eq:H_g} and \ref{eq:R_g} in Eq.~(\ref{eq:rhprima}) we get:
\begin{equation}
R_{g-1}=\frac{28}{3}\left(\frac{3}{2}\right)^{g}-3.
\end{equation}
Now, from the hub of generation $g-1$ the RW needs to reach the pertaining rims and this is first accomplished in a mean time given by:
\begin{eqnarray}\label{eq:hrprima}
H_{g-1}&=&\frac{1}{2(2^{g-1}-1)}\Big[2^{(g-1)-1} \\
\nonumber
&+& \sum_{l=1}^{(g-1)-2}(1+R_l+H_{g-1})+(1+R_{g-1}+H_{g-1})\Big],
\end{eqnarray}
in fact, the RW can jump directly in any of the $2(2^{(g-1)-1})$ rims of generation $(g-1)-1$, either it can reach rims of $g-1$ from which it needs to return to the hub of generation $(g-1)$ in a time $R_{g-1}$ (Eq.~\ref{eq:rhprima}), either it can jump to a rim of the subgraph of generation $g-l$ from which it can return to the hub in $R_l$.

As can be deduced from the previous equations, $H_{g-n}$ and $R_{g-n}$ are deeply connected with each other, such an interplay allows to build up a system of recursive equations. We can in fact write the following general equations
\begin{eqnarray}\label{eq:hrgeneral}
H_{g-n}&=&\frac{1}{2(2^{g-n}-1)}\Big[2^{(g-n)-1}\\
\nonumber
&+& \sum_{l=1}^{(g-n-2)}(1+R_l+H_{g-n})+(1+R_{g-n}+H_{g-n})\Big].
\end{eqnarray}
and
\begin{eqnarray}\label{eq:rhgeneral}
\nonumber
R_{g-n}&=&\frac{1}{g-n}\Big[ (H_{g-n}+1+ R_{g-n})+1\\
&+&\sum_{l=1}^{g-2} (1+H_l+R_{g-n}) \Big].
\end{eqnarray}
Now, by plugging in Eqs.~\ref{eq:rhgeneral} and \ref{eq:hrgeneral} the expressions Eqs.~\ref{eq:R_g} and \ref{eq:H_g}, with some algebra we get
\begin{eqnarray}
R_{g-n}&=&H_{g-n+1}+2+4\left[\left(\frac{3}{2}\right)^{g-n}-1\right], \\
H_{g-n}&=&2R_{g-n+1}+1+\frac{8}{9}\left(\frac{3}{2}\right)^{g-n}.
\end{eqnarray}
By properly combining these equations we have
\begin{equation}
R_{g-n}=R_{g-n+2}-1+\frac{28}{9}\left(\frac{3}{2}\right)^{g-n}
\end{equation}
and using Eq.~\ref{eq:R_g} we have
\begin{equation}
R_{g-n}=4\left(\frac{3}{2}\right)^{g}2^{n/2}-(2^{\frac{4+n}{2}}-1)-\frac{8}{9}\left(\frac{3}{2}\right)^{g-n}.
\end{equation}
Repeating the same calculations for $H_{g-n}$ we obtain similar results, namely:
\begin{equation}
H_{g-n}=4\left(\frac{3}{2}\right)^{g}2^{n/2}-(2^{\frac{4+n}{2}}-3)-\frac{8}{3}\left(\frac{3}{2}\right)^{g-n}.
\end{equation}

Following analogous operations for $n$ odd, we reach the same result provided that $n$ is replaced by $n+1$.We notice that, for a given generation $g$, the calculation of $\tau_g(\mathcal{F})$ involves either $R_{g-2n}$ or $R_{g-2n+1}$ occur according to whether $g$ is even or odd. Hence, in the following, without loss of generality, we focus on the case of $g-n$.

From $H_{g-n}$ and $R_{g-n}$ it is now possible to derive a closed form expression for the trapping time at $\mathcal{F}_g$. 
We now use the partial results found so far to finally calculate $\tau_g(\mathcal{F}_g)$. First of all, let us calculate the first passage time $t_g^H(\mathcal{F}_g)$ on any site in $\mathcal{F}_g$ for a RW started from the main hub; when $g$ is even, $\mathcal{F}_g$ is made up of nodes which correspond to hubs of generation $1$, vice versa, when 
$g$ is odd  $\mathcal{F}_g$ is made up of nodes which correspond to rims of generation $1$.
Let us now focus on $g$ even; we then reach the following expression:
\begin{equation}
t_g^H(\mathcal{F}_g)=\sum_{l=0}^{g-2}R_{g-l}+H_{g-l},
\end{equation}
which can be straightforwardly generalized to the case of a RW started on a hub of generation $g-n$
\begin{equation} \label{tempo_h}
t_{g-n}^H(\mathcal{F}_g)=\sum_{l}^{g-2}(R_{g-l}+H_{g-l}), 
\end{equation}
from which we get
\begin{eqnarray}
 \nonumber
t_{g-n}^H(\mathcal{F}_g)&=& 8 \left( 2^{g/2} -  2^{n/2} \right)  \left[ \left( \frac{3}{2} \right)^g - 1\right] \\
  &-&\frac{32}{5}  \left[ \left( \frac{3}{2} \right)^{g-n} - 1\right] + 2(g- n).
\end{eqnarray}

Let us now calculate the mean time $t_g^R(\mathcal{F}_g)$ taken by a RW started from a rim of generation $g-n$ to first reach any node in $\mathcal{F}_g$; the following relation holds 
 \begin{equation} \label{tempo_h2}
t_{g-n}^R(\mathcal{F}_g)\sum_{l-1}^{g-2}R_{g-l}+\sum_{l=g-n-1}^{g-2}H_{g-l},
\end{equation} 
which is just a truncated version of sum (\ref{tempo_h}), so that we get
\begin{eqnarray}
\nonumber
t_{g-n}^R(\mathcal{F}_g)&=& 4 \left(2 \times 2^{g/2} - 3 \times 2^{n/2} \right) \left[ \left( \frac{3}{2} \right)^g -1 \right] \\
&-& \frac{56}{15} \left ( \frac{3}{2} \right)^{g-n}  +2 (g - n) + \frac{17}{5}.
\end{eqnarray}

In order to simplify the notation we introduce $T_{n}$ which represents the mean time taken by the RW to first reach a site $f$ in $\mathcal{F}_g$ and started from a node at a chemical distance $n$ from $f$. Since we assumed $g$ to be even, when $n$ is even (odd) the starting point is a hub (rim).
By exploiting the intrinsic symmetry of the graph, $T_n$ can be written as the sum of two contributes, according to whether the starting node belongs to the subgraph $\mathcal{G}_{g-1}$ (and therefore the RW has to pass through the main hub before reaching the target) or to $\mathcal{G}_{g-1}' \cup \mathcal{G}_{g-1}''$ (and therefore the RW has to pass through $\mathcal{B}_g$); for subgraphs of smaller generation it holds analogously. The first contribute reads as
\begin{equation}
\sum_{\substack{n=1 \\n \; \mathrm{odd}}}^{g-1}2^{\frac{n-1}{2}}(3^{g-n}-1)(\tau^{H}_{g-n}+T_{g-n+1})+T_{g-n+1},
\end{equation}
in fact the RW has first to reach the main hub of the pertaining subgraph in a mean time $\tau_{g-n}$, from which it reaches $\mathcal{F}_g$ in a time $T_{g-n+1}$. The factor $2^{\frac{n-1}{2}}$ accounts for the number of subgraphs of generation $g-n$ present.
The second contribute reads as
\begin{equation}
4 \sum_{\substack{n=2 \\n \; \mathrm{even}}}^{g-2}2^{\frac{n-2}{2}}(3^{g-n}-2^{g-n})(\tau^{R}_{g-n}+T_{g-n+1})+2^{g-n} T_{g-n+1},
\end{equation}
where we considered RWs starting from nodes such that in order to reach $\mathcal{F}_g$ has first to reach the rims of the pertaining subgraphs or $\mathcal{B}_g$ itself.
We sum together the previous expressions to obtain
\begin{eqnarray}\label{final}
\nonumber
\tau_g(\mathcal{F}_g)&=&\frac{2^{g+1}}{45(3^g - 2^{\frac{g}{2}})} \Big[ 4 \left( \frac{3}{2} \right)^{2g} (45 \times 2^{g/2} - 97)\\ 
\nonumber
&-&   \left( \frac{3}{2} \right)^{g} (45 \times 2^{g/2+2}  -25 g - 527)  -144 \Big],
\end{eqnarray}
for $g$ odd analogous results are expected.


\begin{thebibliography}{}
\bibitem{redner}
S. Redner, \emph{A Guide to First-Passage Processes}, Cambridge University Press, Cambridge (2001)
%
\bibitem{montroll}
E.W. Montroll, J.\ Math.\ Phys.  \textbf{10}, 753 (1969)
%
\bibitem{weiss}
G.H. Weiss, \emph{Aspect and Applications of the Random Walk}, North-Holland, Amsterdam (1994)
%
\bibitem{fisher}
M.E. Fisher, J.\ Stat.\ Phys. \textbf{34}, 667 (1984)
%
\bibitem{chem}
D.-J. Heijis, V.A. Malyshev and J. Knoester, J.\ Chem.\ Phys. \textbf{121}, 4884 (2004).
%
\bibitem{zhang2}
Z. Zhang, W. Xie, S. Zhou, S. Gao and J. Guan, Europhys.\ Lett. \textbf{88}, 100001 (2009).
%
\bibitem{olivier2}
O. B\'{e}nichou, B. Meyer, V. Tejedor,  R. Voituriez, Phys.\ Rev.\ Lett.  \textbf{101}, 130601 (2008).
 %
 \bibitem{condamin}
S. Condamin, O. B\'{e}nichou, V. Tejedor,  R. Voituriez, J. Klafter, Nature  \textbf{450}, 77 (2007).
%
  \bibitem{zhang3}
Z. Zhang, Y. Qi, S. Zhou, W. Xie and J. Guan, Phys.\ Rev.\ E \textbf{79}, 021127 (2009).
%
\bibitem{zhang}
Z. Zhang, Y. Qi, S. Zhou, S. Gao and J. Guan, Phys.\ Rev.\ E \textbf{81}, 016114 (2010).
%
\bibitem{nostro}
E. Agliari, R. Burioni, Phys.\ Rev.\ E  \textbf{80}, 031125 (2009)
%
\bibitem{bollt}
E.M. Bollt, D. ben Avraham, New J.\ Phys.  \textbf{7}, 26 (2005).
%
\bibitem{cantu}
A. Garcia Cant\'{u} and E. Abad, Phys.\ Rev.\ E \textbf{77}, 031121 (2008).
%
\bibitem{cassi}
R. Burioni and D. Cassi, J.\ Phys.\ A \textbf{38}, R45 (2005).
%
\bibitem{agliari}
E. Agliari, Phys.\ Rev.\ E  \textbf{77}, 011128 (2008)
%
\bibitem{olivier}
V. Tejedor, O. B\'{e}nichou, R. Voituriez, Phys.\ Rev.\ E  \textbf{80}, 065104 (2009).
%
\bibitem{barabasi}
A.-L. Barab\'{a}si, E. Ravasz and T. Vicsek, Phys.\ A \textbf{299}, 559 (2001).
%
\bibitem{iguchi}
K. Iguchi and H. Yamada, Phys.\ Rev.\ E. \textbf{71}, 036144
(2005).
%
\bibitem{cinesi_dist}
Z. Zhang, Y. Lin, S. Gao, S. Zhou, J. Guan, J.\ Stat.\ Mech. P10022 (2009)
%
\bibitem{prox}
E. Agliari, R. Burioni, in preparation.
%
\bibitem{watts}
D.J. Watts, H. Strogatz, Nature \textbf{393} 440 (1998)
%
\bibitem{motter}
T. Nishikawa, A.E. Motter, Y.-C. Lai, F.C. Hoppensteadt, Phys.\ Rev.\ Lett. \textbf{91} 014101 (2003)
%
\bibitem{olivier3}
O. B\'{e}nichou, M. Coppey, M. Moreau, P.H. Suet and  R. Voituriez, Europ.\ Phys.\ Lett.  \textbf{70}, 42 (2005).
 (2009)
 %
\bibitem{zhang4}
Z. Zhang, Y. Lin, S. Zhou, B. Wu and J. Guan, New.\ J.\ Phys. \textbf{11}, 103043 (2009).
%
\bibitem{kozak}
J. J. Kozak and V. Balakrishnan, Phys.\ Rev.\ E  \textbf{65}, 021105 (2002).
%


\end{thebibliography}
\end{document}